\documentclass[twocolumn,aps,showpacs]{revtex4}
\usepackage{amsmath}
\usepackage{graphicx}

\newcommand{\project}[1]{|#1\rangle\langle #1|}   
\newcommand{\iter}{\mathrm{i}}

\begin{document}

\title{Quantification of entanglement by means of convergent iterations} 

\author{Jaroslav \v{R}eh\'{a}\v{c}ek}
\author{Zden\v{e}k Hradil}
\affiliation{Department of Optics, Palack\'y University, 772 00 Olomouc, 
Czech Republic}

\pacs{xx.xx}

\begin{abstract}
An iterative procedure is proposed for the calculation of the 
relative entropy of entanglement of a given bipartite quantum state. 
When this state turns out to be non-separable the algorithm
provides the corresponding optimal entanglement witness measurement.
\end{abstract}
\maketitle

Entanglement is an important resource of quantum information processing.
Although there are quantum protocols based on other features of quantum
mechanics such as quantum superposition principle in celebrated 
Deutsch's searching algorithm \cite{deutsch92}, or protocols where the use of 
entanglement  can be advantageous but is not essential, such as 
signaling through depolarizing channels with memory \cite{macchiavello02}, 
most quantum protocols rely on the existence  of non-separable states. For 
practical purposes it is very important to
quantify entanglement generated by realistic laboratory sources and
thus evaluate the potential usefulness of a given realistic 
source for the quantum processing/communication purposes. 

One of the measures of entanglement thoroughly studied over the past decade 
is the relative entropy of entanglement defined as \cite{vedral97}
\begin{equation}\label{entr-entangl}
E(\sigma)=\inf_{\rho\mathrm{\,sep.}} S(\sigma||\rho),
\end{equation} 
where $S$ is the quantum relative entropy,
\begin{equation}
\label{rel-ent}
S(\sigma ||\rho )=\mathrm{Tr}(\sigma \ln \sigma -\sigma \ln \rho ),
\end{equation}
between states $\sigma$ and $\rho$; the infimum in Eq.~(\ref{entr-entangl}) 
is taken over the set of separable states. This functional
is one possible generalization of the classical relative entropy between
two probability distributions \cite{braunstein96} to quantum theory. 
Let us mention that  unlike in the case of entropy this generalization is by 
no means unique.
This quantity can be given a geometrical interpretation as
a quasi-distance between the state whose entanglement we are interested in
and the convex set of separable states. $E$ fulfills most of the requirements
usually imposed on a good entanglement measure and has other good
properties. Most notably, on the set of pure states it coincides
with the Von Neumann reduced entropy \cite{vedral98,plenio00}, and 
is closely related to some other measures of entanglement \cite{henderson00}.
Relative entropy makes also a good entanglement measure for multipartite 
\cite{plenio01} and infinite-dimensional \cite{eisert02} quantum systems.

The analytical form of $E$ is known only for some special sets of states
of high symmetry \cite{vedral97b,audenaert01,vollbrecht01}. 
Generally one has to resort to numerical calculation. 
In a sense  the problem resembles the reconstruction of quantum states  
using the maximum likelihood principle \cite{hradil97,rehacek01}. Here the 
given  input state $\sigma$ plays the role of experimental data; 
once this state is known, the statistics of any possible measurement performed 
on it is available.  The solution can be obtained by means 
of several numerical methods.  The formulation given in \cite{vedral98}
is just an  example  corresponding to an implementation of the downhill 
simplex method.  Its efficiency  strongly depends on the dimensionality of 
the problem.
 
In the procedure proposed here more analytical approach will be adopted.
We will derive a set of extremal equations for $E$ and will show how
to solve them by means of repeated convergent iterations.  
But this is not the only goal. The extremal equations indicate that there is 
a structure of quantum measurement  associated with the extremal solution.
The separable measurement obtained in this way specifies the extremal separable 
state and, significantly, it provides the optimal entanglement witness operator 
revealing the possible entanglement of the input state $\sigma.$

Let us denote $\rho^*$ the separable state having the smallest quantum
relative entropy with respect to $\sigma$. 
Let $f(x,\rho^*,\rho)=S\bigl(\sigma||(1-x)\rho^*+x\rho\bigr)$ be the relative entropy
of a state obtained by moving from $\rho^*$ towards some $\rho$.
We are looking for the global maximum of a convex functional on
the convex set of separable states. Two cases may arise.
When  $\sigma$ is separable the necessary and sufficient condition 
for the maximum of $S$ is that its variations along the paths lying in the 
set of separable states vanish,
\begin{equation}\label{stationary}
\frac{\partial f}{\partial x}(0,\rho^*,\rho)= 0,\quad\forall \rho\quad
\mathrm{separable}.
\end{equation}
When $\sigma$ is entangled $S$ attains its true maximum outside the set of
separable states and we must carry on the maximization on the boundary.
In that case Eq.~(\ref{stationary}) holds only for variations along the 
boundary. 
It is well known that any separable state from the Hilbert space
of dimension $p=d\otimes d$ can be expressed as  a convex sum of (at most) 
$p^2$ projectors on disentangled pure states (Caratheodory's theorem, see
also \cite{vedral98}),
\begin{equation}\label{decomp}
\rho =\sum ^{p^{2}}_{k=1} 
\project{\varphi_k^1}\otimes \project{\varphi_k^2}.
\end{equation}
Here $|\varphi _{k}^{1}\rangle$ and $|\varphi_{k}^{2}\rangle$ are
pure states (not normalized) of the systems $1$ and $2$, respectively.
Now by taking squares of the projectors, 
$\project{\varphi_k^{1,2}}\rightarrow
\bigl(\project{\varphi_k^{1,2}}\bigr)^2$, 
one can remove the boundary and make condition (\ref{stationary}) 
universal. The derivation in Eq.~(\ref{stationary}) can easily be
calculated using an integral representation of the logarithm of a 
positive operator \cite{vedral98}. It reads 
\begin{equation}\label{integral}
\begin{split}
\frac{\partial f}{\partial x}(0,\rho^*,\rho)&=
\int_0^{\infty}\mathrm{Tr}
\bigl((\rho^*+t)^{-1}\sigma(\rho^*+t)^{-1}\delta\rho\bigr)\mathrm{d}t\\
&=\mathrm{Tr}A\delta\rho,
\end{split}
\end{equation}
where we denoted $(1-x)\rho^*+x\rho=\rho^*+\delta\rho$, and
operator $A$ has the following matrix elements in the eigenbasis 
$\{|\lambda_n\rangle\}$ of $\rho^*$
\footnote{When \( m=n \) or \( \lambda _{n}=\lambda _{m}\)
the corresponding coefficient should be replaced the limit  
value of \( \lambda _{n}^{-1}. \)},  
\begin{equation}\label{A}
\langle \lambda_m|A|\lambda_n\rangle=\frac{\log\lambda_{n}-\log\lambda_{m}}
{\lambda_{n}-\lambda_{m}}\langle\lambda_m|\sigma|\lambda_n\rangle.
\end{equation}
Its meaning will be discussed later.
The right hand side of Eq.~(\ref{integral}) should vanish for all $\delta\rho$ 
preserving the form (\ref{decomp}) of $\rho^*+\delta\rho$.
There are $p^2$ such basic variations:
$\varphi_k^1\otimes\varphi_k^2\rightarrow 
(\varphi_k^1+\delta\varphi_k^1)\otimes(\varphi_k^2+\delta\varphi_k^2)$.
To make sure that the only constraint $\mathrm{Tr}(\rho^*+\delta\rho)=1$ is 
obeyed we will use a Lagrange multiplier $\lambda$.

Getting all things together we arrive at the extremal equations 
that read
\begin{equation}\label{raw-extr}
\begin{split}
R_k^1\project{\varphi_k^1}&=\lambda\project{\varphi_k^1},\\
R_k^2\project{\varphi_k^2}&=\lambda\project{\varphi_k^2},
\end{split}\quad 
k=1\ldots p^2,
\end{equation}
where 
\begin{equation}
R_{k}^{1}=\mathrm{Tr}_{2}\bigl(A\overline{\project{\varphi_k^2}}\bigr),
\quad 
R_{k}^{2}=\mathrm{Tr}_{1}\bigl(A\overline{\project{\varphi_k^1}}\bigr).
\end{equation}
Bars denote projectors normalized to unity. 
Multiplying the first row of Eq.~(\ref{raw-extr}) by 
$\project{\varphi_k^2}$, the second by $\project{\varphi_k^1}$, 
and summing them separately over $k$ we find that 
$\lambda=\mathrm{Tr}A\rho^*=1$.
Using this and modifying further the necessary condition 
we get the main formal result of this paper: The state having 
the smallest quantum relative entropy with respect to a given state 
$\sigma$ satisfies the following $2p^2$ equations,  
\begin{equation}\label{r-rho-r}
\begin{split}
R_k^1 \project{\varphi_k^1} R_k^1 &= \project{\varphi_k^1},\\
R_k^2 \project{\varphi_k^2} R_k^2 &= \project{\varphi_k^2},
\end{split}\quad
k=1\ldots p^2.
\end{equation}
Unfortunately, solving such highly nonlinear operator equations by 
analytical means for anything but most trivial states seems to be  out
of question. One has to turn to numerics. We suggest to solve 
Eqs.~(\ref{r-rho-r}) by repeated iterations starting from some randomly
chosen separable $\rho$. Let us note that the iterative  procedure based
on Eqs.~(\ref{r-rho-r}) belongs to the family of gradient-type
algorithms of the form  
$x_k^{i+1}=\bigl(\partial S(\mathbf{x}^i)/\partial x_k\bigr) x_k^i 
\bigl(\partial S(\mathbf{x}^i)/\partial x_k\bigr) $.  
Algorithms of this type are known to behave well;
some of them were even proven to converge monotonically
\cite{byrne93}. 
They have found important applications in various 
optimizations and inverse problems.
In our case we observed that the step generated by operators 
$R_k^{1,2}$ in  Eqs.~(\ref{r-rho-r}) was often too large --- 
rather than   converging to the stationary point the algorithm 
would oscillate or diverge. If this happens the length of the step 
can be made smaller by mixing the operators $R_k^{1,2}$ with the unity 
operator: 
\begin{equation}\label{regular}
R_k^{1,2}\rightarrow 
(\openone+\textstyle\frac{1}{2}\alpha R_k^{1,2})/(1+\textstyle\frac{1}{2}
\alpha).
\end{equation} 
Indeed, when $\alpha$ is sufficiently small the 
algorithm converges monotonically. This can be seen by considering 
an infinitesimal step with $\alpha\ll 1$. It is convenient 
to split one iteration of Eqs.~(\ref{r-rho-r}) into two subsequent steps
corresponding to the two rows of Eqs.~(\ref{r-rho-r}) (projectors
of only one of the subsystems are updated at a time).
The two steps are completely symmetrical, so we will consider
an infinitesimal iteration on, say, the projectors 
$\project{\varphi_k^{1\iter}}$ of the first system obtained after the i-th 
iteration. We want to show that after one such step the quantum relative 
entropy is never increased,
$S(\sigma||\rho^{\iter+1})\le S(\sigma||\rho^\iter)$.
Using Eq.~(\ref{regular}) in Eqs.~(\ref{r-rho-r}) we get to the first order 
in $\alpha$,
\begin{equation}
\rho^{\iter+1}=(1-\alpha)\rho^\iter+\alpha\tilde\rho,
\end{equation}
where $\tilde\rho=\textstyle\frac{1}{2}\sum_k \bigl(R_k^{1\iter}
\project{\varphi_k^{1\iter}}+\project{\varphi_k^{1\iter}} R_k^{1\iter}\bigr)
\otimes\project{\varphi_k^{2\iter}}$,
and thus 
\begin{equation}
S(\sigma||\rho^{\iter+1})-S(\sigma||\rho^\iter) \propto 
\frac{\partial f(0,\rho^\iter,\tilde\rho)}{\partial\alpha}=
1-\mathrm{Tr}A^\iter\tilde\rho.
\end{equation}
It remains to show that $\mathrm{Tr}A^\iter\tilde\rho\ge 1$. Let us denote 
$\lambda^\iter_k=\langle\varphi_k^{1i}|\varphi_k^{1\iter}\rangle
\langle\varphi_k^{2\iter}|\varphi_k^{2\iter}\rangle$. Notice that 
$\sum_k\lambda_k^\iter=1$ 
by the normalization of $\rho^\iter$. Then by using the Swartz inequality
and the concavity of the square function we obtain
\begin{equation}
\begin{split}
\mathrm{Tr}A^\iter\tilde\rho&=\sum_k\lambda_k^\iter\mathrm{Tr}\bigl(R_k^{1\iter}
\overline{\project{\varphi_k^{1\iter}}}R_k^{1\iter}\bigr)\\
&\ge \sum_k\lambda_k^\iter\Bigl[\mathrm{Tr}\bigl(R_k^{1\iter}\,
\overline{\project{\varphi_k^{1\iter}}}\bigr)\Bigr]^2\\
&\ge \Bigl[\sum_k\lambda_k^\iter\mathrm{Tr}\bigl(R_k^{1\iter}\,
\overline{\project{\varphi_k^{1\iter}}}\bigr)\Bigr]^2\\
&=\bigl(\mathrm{Tr}A^\iter\rho^\iter\bigr)^2=1,
\end{split}
\end{equation}
which completes our proof. 
This means that with a sufficient amount of regularization 
(\ref{regular}) the algorithm (\ref{r-rho-r}) converges monotonically. 
In practice, the parameter $\alpha$ need not be very small.
In $2\otimes 2$ and $4\otimes 4$ dimensional problems we tried 
monotonic convergence was observed even with $\alpha$ of the order of 
unity.
   
\begin{figure}
\centerline{
\includegraphics[width=0.6\columnwidth]{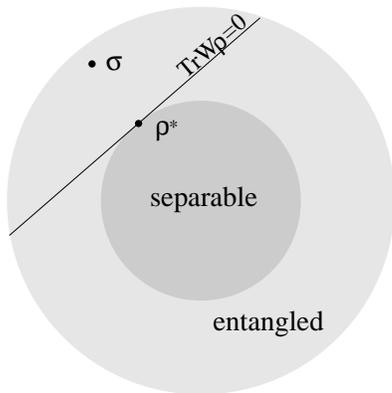}
}
\caption{Mutual relationship between an entangled state $\sigma$, the separable
state $\rho^*$ closest to it in the sense of quantum relative entropy,
and the entangled states detected by the witness operator $W$.
\label{fig:geom}}
\end{figure}

Now let us go back to the extremal equations (\ref{r-rho-r}).
The left hand sides are generated by the operator $A$, which
depends on $\sigma$ through Eq.~(\ref{A}). Let us assume for 
now that $\sigma$ is an entangled state. Then $A$ represents
the gradient of the quantum relative entropy 
$S(\sigma||\rho)$ at $\rho^*$---the separable state 
closest to $\sigma$. Loosely speaking, the states giving the same
expectation, $\mathrm{Tr}A(\sigma)\rho=\mathrm{const.}$, form hyper-planes
that are perpendicular to the line connecting $\sigma$ and $\rho^*$.
Since $\mathrm{TrA(\sigma)\rho^*=1}$ and $\rho^*$ lies at the
boundary of the set of separable states, the conjecture is
that the operator $A$ is up to a shift of its spectrum 
a witness operator \cite{werner89,terhal00} detecting the entanglement of 
$\sigma$. In the following we will show that this is indeed the case,
and that the operator 
\begin{equation}\label{oper-w}
W(\sigma)=\openone-A(\sigma)
\end{equation} 
is indeed the optimal witness of the entanglement of $\sigma$.
The mutual relationship of $\sigma$, $\rho^*$, and the states detected
by $W$ is shown in Fig.~\ref{fig:geom}.

First we will show that $\mathrm{Tr}A\rho\le 1$ if $\rho$ is separable.
To this end let us note that 
\begin{equation}\label{proof1}
\begin{split}
1-\mathrm{Tr}A\rho&=\frac{\partial f}{\partial x}(0,\rho^*,\rho)\\
&=\lim_{x\rightarrow 0}
\frac{S\bigl(\sigma||(1-x)\rho^*+x\rho\bigr)-S(\sigma||\rho^*)}{x}.
\end{split}
\end{equation}   
Now, since both $\rho$ and $\rho^*$ are separable states,
so is their convex combination $(1-x)\rho^*+x\rho$. But $\rho^*$
minimizes $S(\sigma||\rho)$ over the set of separable states.
Therefore, $S\bigl(\sigma||(1-x)\rho^*+x\rho\bigr)-S(\sigma||\rho^*)\ge 1$.
This holds for all $x$ so we have,
\begin{equation}\label{proof12}
\mathrm{Tr}{W(\sigma)\rho}=1-\mathrm{Tr}{A(\sigma)\rho}\ge 0,\quad \forall 
\rho \mathrm{\,separable}
\end{equation}
This already means that $W$ is an entanglement witness operator.
To show that $W$ detects $\sigma$ we will again make use of 
Eq.~(\ref{proof1})
with $\rho$ now being substituted by the entangled state $\sigma$.
Now, because of convexity of $S$, 
\begin{equation}\label{proof2}
\frac{S\bigl(\sigma||(1-x)\rho^*+x\sigma\bigr)-S(\sigma||\rho^*)}{x}
\le -S(\sigma||\rho^*)< 0.
\end{equation}
The last inequality follows from the assumed non-separability of  
$\sigma$. Eq.~(\ref{proof2}) also holds for any $x$ so we obtain
\begin{equation}\label{proof22}
\mathrm{Tr}{W(\sigma)\sigma}=1-\mathrm{Tr}{A(\sigma)\sigma}<0, \quad\forall
\sigma \mathrm{\,entangled}
\end{equation}
which we set out to prove.

\begin{figure}
\centerline{
\includegraphics[angle=270,width=\columnwidth]{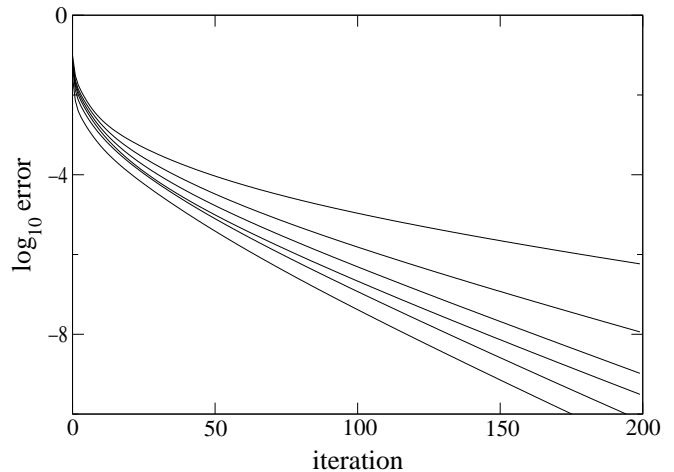}
}
\caption{Test of separability. It is shown how the calculated  relative entropy
of entanglement of several randomly generated separable states approaches 
zero in the course of iterating. The ordinate is labeled by the precision
in decimal digits. 
\label{fig:separ}}
\end{figure}

Possible applications of our algorithm are twofold:
First, it can be used for checking whether a given state is separable 
or not. Second, it can be used for quantifying the amount of entanglement
the state contains. 
As a test of separability we tested the algorithm on many randomly 
generated separable and NPT states of dimensions $2\otimes 2$ and 
$4\otimes 4$; typical results are summarized in Fig.~\ref{fig:separ}.

Recently, another numerical test of separability has been proposed 
\cite{doherty02} consisting of a hierarchy of gradually more and more 
complex separability criteria that can be formulated as separate 
problems of  the linear optimization theory. The algorithm we propose is 
much more simple. There is just one set of equations to
be solved by repeated iterations and after that one finds  
not only whether the input state is entangled but also how much.

\begin{figure}
\centerline{
\includegraphics[width=\columnwidth]{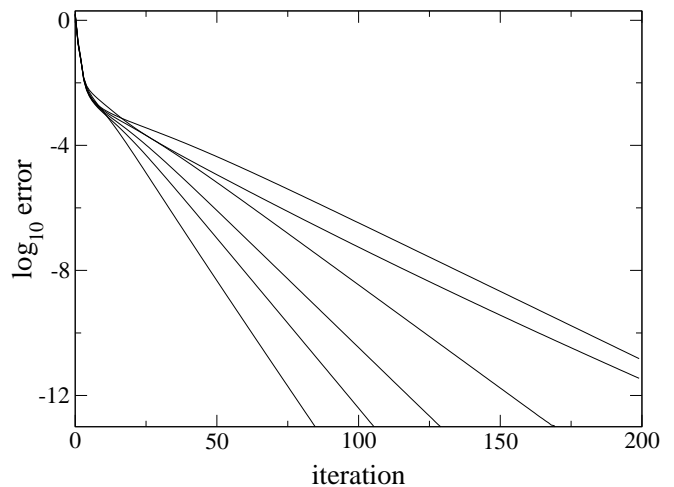}
}
\caption{The accuracy of the calculated relative entropy of entanglement
is shown after a given number of iterations for six Werner states
with $f\in[-0.05,-0.8]$. The ordinate is labeled by the precision in
decimal digits.
\label{fig:werner}}
\end{figure}

Unfortunately, explicit formulas for the quantum relative entropy  
are known only in very few cases.  One such exception is the family 
of Werner states defined as follows,
\begin{equation}\label{werner}
\rho_{\textrm{w}}=\frac{d-f}{d(d^2-1)}\openone+
\frac{fd-1}{d(d^2-1)}F,
\end{equation}
where $F$ is the flip operator $F(\psi^1\otimes\psi^2)
=\psi^1\otimes\psi^2$ and $f=\mathrm{Tr}\rho_{\mathrm{w}}\!F
\in[-1,1]$ is a parameter. 
Fig.~\ref{fig:werner} shows the performance of our algorithm
for several entangled Werner states of dimension $4\otimes 4$.
It is worth mentioning that the optimal entanglement 
witness  $W$ for the detection of Werner states generated by the operator 
$A$ Eq.~(\ref{A}) is simply $W=1-2|\Psi_{-}\rangle\langle\Psi_-|$,
where $\Psi_{-}$ is the singlet state. The expectation value of $W$
is then a renormalized singlet fraction. 

The curves appearing in Figs.~\ref{fig:separ} and 
\ref{fig:werner} suggest that the convergence  of the proposed
algorithm is faster than polynomial but slower than exponential. 
In some cases such as that shown in Fig.~\ref{fig:werner} the convergence
is nearly exponentially fast. This is just a qualitative statement since 
we did not attempt to do any optimization of the length of the iteration 
step. In all probability such optimization would result in further
speedup compared to our examples given in Figs.~\ref{fig:separ} and 
\ref{fig:werner}. 

In conclusion, we derived a convergent iterative algorithm for
the calculation of the relative entropy of entanglement.
It can be used for checking whether a given input state is entangled. 
If it is, the algorithm calculates its relative entropy of 
entanglement, finds its closest separable state, and provides the 
optimal entanglement witness measurement. 

\begin{acknowledgments}
This work was supported by the projects J14/98 and LN00A015 of the Czech 
Ministry of Education.
\end{acknowledgments}

\end{document}